\documentclass[10pt,a4paper]{article}
\usepackage{amsmath}
\usepackage{amssymb}
\usepackage{epsfig}

\newtheorem{thm}{Theorem}
\newtheorem{col}{Corollary}

\def\sn{\mathop{\rm sn}\nolimits}
\def\cn{\mathop{\rm cn}\nolimits}
\def\dn{\mathop{\rm dn}\nolimits}

\title{
Mechanical Systems with Poincar\'e Invariance
}

\author{H. W. Braden\thanks{E-mail:hwb@ed.ac.uk}\ \ and
J.G.B. Byatt-Smith\thanks{E-mail:Byatt@ed.ac.uk}\\
\normalsize
\em Department of Mathematics and Statistics,\\
\normalsize
\em The University of Edinburgh, \\
\normalsize
\em Edinburgh, UK \\
}

\date{September, 2001}

\begin{document}

\renewcommand{\thepage}{}
\begin{titlepage}

\maketitle
\vskip-9.5cm
\hskip10.4cm
%MS-01-00?
%\vskip.2cm
%\hskip10.4cm
%\sf solv-int/9804082 \rm
\vskip8.8cm

\begin{abstract}
Some years ago Ruijsenaars and Schneider initiated  the study of
mechanical systems exhibiting an action of the Poincar\'e algebra.
The systems they discovered were far richer: their models were
actually integrable and possessed a natural quantum version.
We follow this early work finding and classifying mechanical
systems with such an action.
New solutions are found together with a new class of models exhibiting an
action of the Galilean algebra. These are related to the functional
identities underlying the various Hirzebruch genera.
The quantum mechanics is also discussed.
\end{abstract}

\begin{flushleft}
\textbf{2000 AMS Subject Classification}: Primary
39B32 %Difference and Functional equations in complex variables
%39B22 %Difference and Functional equations in real variables
30D05  %Functional Equations in Complex Domain
33E05 %Elliptic Functions and Elliptic Integrals
\end{flushleft}

\begin{flushleft}
\textbf{Key Words}: Mechanics, Integrability, Functional Equations
\end{flushleft}

\vfill
\end{titlepage}
\renewcommand{\thepage}{\arabic{page}}

\section{Introduction}
In this letter we shall introduce new mechanical models obeying the
Poincar\'e algebra
\begin{equation}
\{H,B\}=P ,\qquad \{P,B\}=H,\qquad \{H,P\}=0.
\label{poincare}
\end{equation}
Here $H$ will be the Hamiltonian of the system generating time-translations,
$P$ is a space-translation generator and $B$ the generator of boosts.
The investigation of such models was initiated by Ruijsenaars and
Schneider \cite{RS}.
The models they discovered were found to posses other nice features: they were
in fact integrable and a quantum version of them naturally existed.
We shall discuss these further features of our models below.

Ruijsenaars and Schneider \cite{RS} began with the ansatz
$$
H=\sum_{j=1}\sp{n}\cosh p_j\, \prod_{k\ne j}f(x_j-x_k)
,\qquad
P=\sum_{j=1}\sp{n}\sinh p_j\, \prod_{k\ne j}f(x_j-x_k)
$$
and
$$B=\sum_{j=1}\sp{n}x_j .  $$
With this ansatz and the canonical Poisson bracket $\{p_i,x_j\}=\delta_{ij}$
the first two Poisson brackets of (\ref{poincare}) involving the boost
operator $B$ are automatically satisfied.
The remaining Poisson bracket is then
\begin{multline*}
\{H,P\}=-\sum_{j=1}\sp{n}\partial_j \prod_{k\ne j}f\sp2(x_j-x_k) \\
-\frac{1}{2}\sum_{j\ne k}\cosh(p_j-p_k)\, \prod_{l\ne j}f(x_j-x_l)
\prod_{m\ne k}f(x_k-x_m)\Big(\partial_j\ln f(x_k-x_j)+\partial_k\ln f(x_j-x_k)
\Big)
\end{multline*}
and for the independent terms proportional to $\cosh(p_j-p_k)$ to vanish
we require that $f'(x)/f(x)$ be odd.
This entails that $f(x)$ is either even or
odd\footnote{Ruijsenaars and Schneider assume $f(x)=f(-x)$.}
and in either case $f\sp2(x)$ is even.
%$$\{H,B\}=P ,\qquad \{P,B\}=H$$
Supposing that $f(x)$ is so constrained, then the final Poisson bracket is
equivalent to the functional equation
\begin{equation}
\{H,P\}=0 \Longleftrightarrow
\sum_{j=1}\sp{n}\partial_j \prod_{k\ne j}f\sp2(x_j-x_k)=0.
\label{functional}
\end{equation}
For $n=3$ this equation may be written in the form
\begin{equation}
\begin{vmatrix}
1              & 1              & 1              \\
F(x)           & F(y)           & F(z)           \\
F\sp{\prime}(x)& F\sp{\prime}(y)& F\sp{\prime}(z)\\
\end{vmatrix}
=0,\quad\quad  x+y+z=0,
\label{detdiff}
\end{equation}
where $F(x)=f\sp2(x)$. Ruijsenaars and Schneider \cite{RS}
showed that $F(x)=\wp(x)+c$
satisfies (\ref{detdiff}) and further satisfies (\ref{functional}) for all $n$.
This same functional equation (without assumptions on the parity of the
function $F(x)$) has arisen in several settings related to integrable systems.
It arises when characterising quantum mechanical potentials whose ground
state wavefunction (of a given form) is factorisable \cite{cal, suth, Gu}.
More recently it has been shown \cite{Bra} to characterise the Calogero-Moser
system \cite{vDV}, which is a scaling limit of the Ruijsenaars-Schneider system.
The analytic solutions to (\ref{detdiff}) were characterised by Buchstaber
and Perelomov \cite{bp} while more recently a somewhat stronger result with
considerably simpler proof was obtained by the authors \cite{BBS}. One has
\begin{thm}
Let F be a three-times differentiable function satisfying the functional
equation (\ref{detdiff}).
Then, up to the manifest invariance
$ F(x)\rightarrow \alpha F(\delta x)+\beta$,
the solutions of (\ref{detdiff}) are one of $F(x)=\wp(x+d)$,
$F(x)=e\sp{x}$ or $F(x)=x$.
Here $\wp$ is the Weierstrass $\wp$-function and
 $3 d$ is a lattice point of the $\wp$-function.
\end{thm}

Thus the even solutions of (\ref{detdiff}) are precisely those obtained by
Ruijsenaars and Schneider.  As an aside we remark that the delta function
potential $a\delta \left( x\right) $
of many-body quantum mechanics on the line, which has a factorisable
ground-state wavefunction, can be viewed as the $\alpha \rightarrow 0$
limit of $-{b}/{\alpha \sinh^{2}\left( -x/\alpha +\pi i/3\right) } $
with $\pi a\alpha =6b$. Thus all of the known quantum
mechanical problems with factorisable ground-state wavefunction
are included in (\ref{detdiff}).
There appear deep connections between functional equations and integrable
systems \cite{Ca2, BCb, inoz, Inoz, BKr, DFS, BB1, BB2, Gur1, Gur2, Gur3, Gur4}.

At this stage all we know is that the $n=3$ solution of Ruijsenaars and
Schneider to (\ref{functional}) in fact yields a solution for all $n$.
The general even solution of (\ref{functional}) has not been given. Our
first new result is
\begin{thm}
The  general even solution of (\ref{functional}) amongst the class of
meromorphic functions whose only singularities on the Real axis are
either a double pole at the origin, or double poles at $n p$ ($p$
real, $n\in \mathbb Z$) is:
\newline a) for all odd $n$ given by the solution of Ruijsenaars and
Schneider while\newline
b) for even $n\ge4$ there
are in addition to the Ruijsenaars-Schneider solutions the following:
\begin{equation*}
\begin{split}
F_1(z)&=\sqrt{(\wp(z)-e_2)(\wp(z)-e_3)}=
\frac{\sigma_2(z)\sigma_3(z)}{\sigma\sp2(z)} \\&=
\frac{\theta_3(v)\theta_4(v)}{\theta_1\sp2(v)}
\frac{\theta_1\sp{\prime2}(0)}{4\omega^2\theta_3(0)\theta_4(0)}
= b\frac{\dn(u)}{\sn\sp2(u)} \\
F_2(z)&=\sqrt{(\wp(z)-e_1)(\wp(z)-e_3)}=
\frac{\sigma_1(z)\sigma_3(z)}{\sigma\sp2(z)} \\&=
\frac{\theta_2(v)\theta_4(v)}{\theta_1\sp2(v)}
\frac{\theta_1\sp{\prime2}(0)}{4\omega^2\theta_2(0)\theta_4(0)}
=b\frac{\cn(u)}{\sn\sp2(u)}\\
F_3(z)&=\sqrt{(\wp(z)-e_1)(\wp(z)-e_2)}=
\frac{\sigma_1(z)\sigma_2(z)}{\sigma\sp2(z)} \\ &=
\frac{\theta_2(v)\theta_3(v)}{\theta_1\sp2(v)}
\frac{\theta_1\sp{\prime2}(0)}{4\omega^2\theta_2(0)\theta_3(0)}
=b\frac{\cn(u)\dn(u)}{\sn\sp2(u)}
\end{split}
\end{equation*}
\end{thm}
Here
$$\sigma_\alpha(z)=
\frac{\sigma(z+\omega_\alpha)}{\sigma(\omega_\alpha)}
e\sp{-z\zeta(\omega_\alpha)},
\qquad u=\sqrt{e_1-e_3}\,z ,
\qquad v=\frac{z}{2\omega},
\qquad b={e_1-e_3}
$$
with $\omega_1=\omega$, $\omega_2=-\omega-\omega'$ and $\omega_3=\omega'$,
and we have given representations in terms of the Weierstrass elliptic
functions, theta functions and the Jacobi elliptic functions \cite{WW}.
The functions $F_k(z)$ are elliptic functions whose
periodicities and (double) poles are as follows:
$$
\begin{array}{ll}
F_1(u)= F_1(u+2K)=F_1(u+4iK')&u=0, 2iK', \\
F_2(u)= F_2(u+4K)=F_u(u+2K+2iK')=F_2(u+4iK')&u=0, 2K,\\
F_3(u)= F_3(u+2iK')=F_3(u+4K)&u=0, 2K.
\end{array}
$$
These functions satisfy $( {d\sp2}/{dx\sp2}-6/\sn\sp2(x) )F_i=
-\lambda_i F_i$ ($\lambda_i=4+k\sp2, 1+4 k\sp2, 1+k\sp2$
respectively) and $F_i(u+iK')$ are $g=2$ Lam\'e polynomials. (The remaining two
$g=2$ Lam\'e polynomials are of Ruijsenaars-Schneider form with similarly
shifted argument.)
The functions $F_k(z)$ have the expansion $F_k(z)=1/z^2 +e_k/2 +O(z^2)$,
and if $P$ denotes the other pole in the fundamental region determined by
its periodicity we have
\begin{equation}
F_k(z+P)=-F_k(z).
\label{periodicity}
\end{equation}
For appropriate ranges of $z$ the solutions are real.
Their degenerations yield all the even solutions with only a double pole at
$x=0$ on the real axis.  These degenerations may in fact coincide with
the degenerations of the Ruijsenaars-Schneider solution.

One can straightforwardly verify these new solutions do in fact satisfy
(\ref{functional}) for even $n$ but new techniques needed to be developed to
show we have exhausted the solutions.
(These will be presented elsewhere \cite{BBS2}.)
One method of verifying these are solutions is as follows.
Let $F$ be any one of $F_1$, $F_2$ or $F_3$.
Setting $x_{n}=x_{n-1}+\epsilon+\alpha P$, with $\alpha=0,1$ in
(\ref{functional}) yields
\begin{multline*}
\sum_{j=1}\sp{n}\partial_j \prod_{k\ne j}\sp{n}F(x_j-x_k)
=\sum_{j=1}\sp{n-2}\partial_j\Big(F(x_j-x_{n-1})F(x_j-x_{n})
\prod_{k\ne j}\sp{n-2}F(x_j-x_k)\Big)\\
+\partial_{n-1}\Big(F(x_{n-1}-x_{n})\prod_{k=1}\sp{n-2}F(x_{n-1}-x_k)\Big)
+\partial_{n}\Big(F(x_{n}-x_{n-1})\prod_{k=1}\sp{n-2}F(x_{n}-x_k)\Big)\\
=\sum_{j=1}\sp{n-2}\partial_j\Big( (-1)\sp{\alpha}
F(x_j-x_{n-1})F(x_j-x_{n-1}-\epsilon)\prod_{k\ne j}\sp{n-2}F(x_j-x_k)\Big)\\
+(-1)\sp{\alpha}\Big[F'(-\epsilon)\prod_{k=1}\sp{n-2}F(x_{n-1}-x_k)+
F(\epsilon)\partial_{n-1}\prod_{k=1}\sp{n-2}F(x_{n-1}-x_k)\Big]+\\
(-1)\sp{\alpha(n-1)}
\Big[F'(\epsilon)\prod_{k=1}\sp{n-2}F(x_{n-1}+\epsilon-x_k)+
F(\epsilon)\partial_{n-1}\prod_{k=1}\sp{n-2}F(x_{n-1}+\epsilon-x_k)\Big]
\end{multline*}
For both $\alpha=0,1$ there are pole terms $\epsilon\sp{-3}$,
$\epsilon\sp{-2}$, $\epsilon\sp{-1}$ as $\epsilon\rightarrow0$.
When $\alpha=0$ it is easy to see these poles vanish.
For example the $\epsilon\sp{-3}$ term vanishes simply from the the oddness of
$F'(\epsilon)$. When $\alpha=1$ (corresponding to the pole $P$ of $F(x)$)
we see we require an even number of particles.
(There are other poles in this expression of the form $\epsilon=x_{n-1}-x_s+
\alpha P$, but these correspond to $x_{n}=x_s+\epsilon+\alpha P$ and
vanish by symmetry and the above arguments.) Viewed as a function of $x_n$
we have a doubly elliptic function with no poles, thus for $n$ even
$\sum_{j=1}\sp{n} \partial_j \prod_{k\ne j} F(x_j-x_k)$ is independent of
$x_n$. By symmetry this is a constant.
The value of this constant may be established as zero by the following
argument. Set $x_j=j\mu$, where $\mu$ is any number such that
$\pm s\mu$ ($s\in \{1,\ldots,n-1\}$) is not a pole of $F(x)$. Consider the
pairing $j\leftrightarrow n-j$. With our choice of $x_j$ then
$F(x_j-x_k)=F(-[x_{n-j}-x_{n-k}])=F(x_{n-j}-x_{n-k})$ while
$F'(x_j-x_k)=-F'(x_{n-j}-x_{n-k})$. Using this involution we may show the terms
appearing in the constant sum cancel pairwise:
\begin{equation*}
\begin{split}
\sum_{j=1}\sp{n} \partial_j  & \prod_{k\ne j} F(x_j-x_k)=
\sum_{j\ne k} F'(x_j-x_k)\prod_{l\ne j,k}F(x_j-x_l)\\ &=
\frac{1}{2}\sum_{j\ne k} \Big(F'(x_j-x_k)+F'(x_{n-j}-x_{n-k})\Big)
\prod_{l\ne j,k}F(x_j-x_l)=0.
\end{split}
\end{equation*}
Thus we obtain $0=\sum_{j=1}\sp{n} \partial_j \prod_{k\ne j} F(x_j-x_k)$
and our functions satisfy the equation.

\section{Integrability}
The models discovered by Ruijsenaars and Schneider not only exhibited
an action of the Poincar\'e algebra but were completely integrable as well
and we will discuss this aspect of our models.
Throughout we henceforth assume that $f$ is either even or odd as discussed
earlier.

Following \cite{RS} we introduce the light-cone quantities
\begin{equation}
S_{\pm k} =\sum_{\substack{ I\subseteq \{1,2,\ldots, n\}\\ \\ |I|=k} }
\exp\left(\pm {\sum_{i\in I} p_i}\right) \,
\prod_{\substack{ i\in I\\ \\ j\not\in I}} f(x_i-x_j).
\label{lccons}
\end{equation}
Then $H=(S_1+S_{-1})/2$ and $P=(S_1-S_{-1})/2$. We investigate the
Poisson commutativity of these $S_k$'s.

Let $I,J\subseteq \{1,2,\ldots, n\}$ and define the sets
\begin{equation*}
A=I\backslash J ,\qquad B=J\backslash I,\qquad C=I\cap J, \qquad
D=\overline{I\cup J}.
\end{equation*}
\begin{center}
\epsfig{file=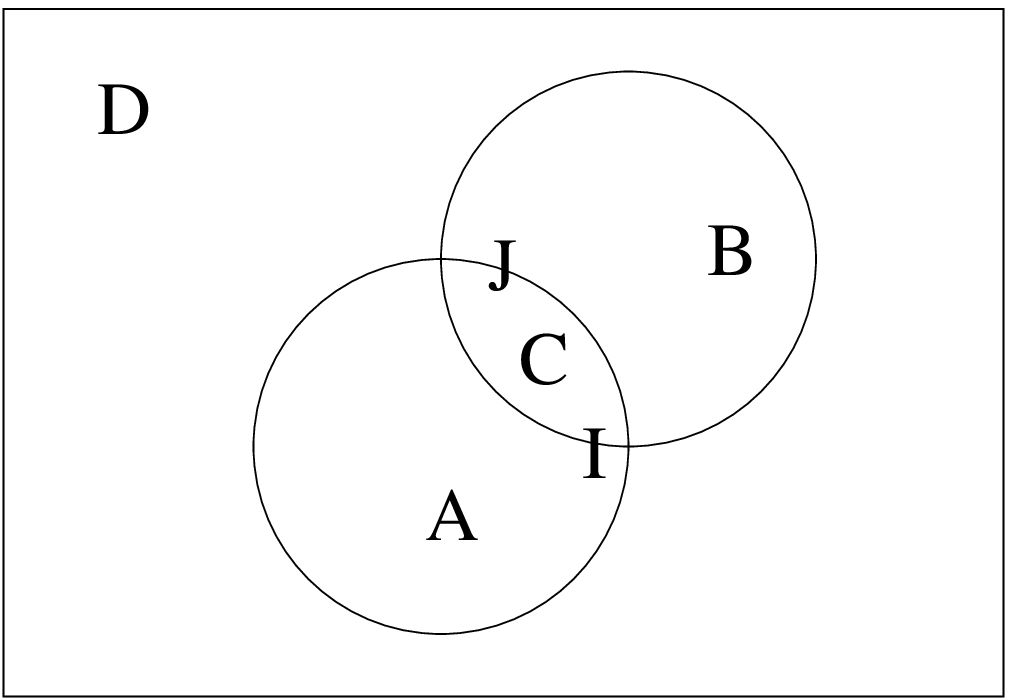, width=2.0in,angle=0}
\end{center}
Then $I=A\cup C$ and so on. Introduce the notation \cite{RS}
\begin{eqnarray*}
(J) &=&\prod\limits_{\substack{ l\in J \\ m\not\in J}}f(x_{l}-x_{m})
=\prod\limits_{\substack{ l\in J \\ m\in A}}f(x_{l}-x_{m})
 \prod_{\substack{ l\in J \\  n\in D}}f(x_{l}-x_{n}) \\
&=&\prod\limits_{\substack{ l\in C \\ m\in A}}f(x_{l}-x_{m})
   \prod\limits_{\substack{ l\in B \\ p\in A}}f(x_{l}-x_{p})
   \prod\limits_{\substack{ l\in B \\ r\in D}}f(x_{l}-x_{r})
   \prod\limits_{\substack{ l\in C \\ s\in D}}f(x_{l}-x_{s})
\end{eqnarray*}
and for disjoint sets $A,B$ set
$$(AB)=\prod\limits_{\substack{ a\in A \\ b\in B}}f(x_{a}-x_{b})=
(\pm1)\sp{|A||B|}(BA)$$
(using the evenness/oddness of $f$).
Thus $(I)=(AB)(AD)(CB)(CD)$.
Also set
\begin{equation*}
e\sp{\pm\theta_J}=e\sp{\pm\sum_{j\in J}p_j}.
\end{equation*}
Then the light-cone quantity $S_k$ becomes
$$S_{\pm k} =\sum_{\substack{ J\subseteq \{1,2,\ldots, n\}\\ \\ |J|=k} }
e\sp{\pm\theta_J}\, (J).$$
Now clearly $\{S_k,S_{\pm n}\}=0$ and $S_{k-n}=S_k\, S_{-n}$
for all $k$, thus we only need focus on $k\in \{1,2,\ldots, n\}$.
To calculate the Poisson bracket $\{S_k,S_l\}$ we must evaluate
expressions of the form
$$
\left\{ e^{\theta _{I}}(I),e^{\theta _{J}}(J)\right\} =
e^{\theta_{I}+\theta_{J}}\left[ (I)\partial_{I}(J)-(J)\partial_{J}(I)\right]
$$
where $\partial_I=\sum_{i\in I}\partial_i$.
Now
\begin{equation*}
\partial_{I}(J)=(J)\left[ -\sum\limits_{\substack{ l\in B \\ p\in A}}\frac{%
f'\left( x_{l}-x_{p}\right) }{f(x_{l}-x_{p})}+\sum\limits_{\substack{ %
l\in C \\ p\in D}}\frac{f'\left( x_{l}-x_{p}\right) }{f(x_{l}-x_{p})}%
\right]
\end{equation*}
whence
\begin{eqnarray*}
\left\{ e^{\theta _{I}}(I),e^{\theta _{J}}(J)\right\}  &=&e^{\theta
_{I}+\theta _{J}}(I)(J)\sum\limits_{\substack{ l\in B \\ p\in A}}\left[
-\frac{f'\left( x_{l}-x_{p}\right) }{f(x_{l}-x_{p})}+\frac{f'\left(
x_{p}-x_{l}\right) }{f(x_{p}-x_{l})}\right]  \\
&=&e^{\theta _{I}+\theta _{J}}(I)(J)\frac{2}{(AB)}\partial _{A}(AB) \\
&=&
(\pm)\sp{(|A|+|B|)|C|+|A||B|}
e^{\theta_{I}+\theta_{J}}(A\cup B)(CD)^{2}\partial_{A}\left( AB\right)^{2}\\
&=&
(\pm)\sp{(|A|+|B|)|C|+|A||B|}
e^{\theta_{A\cup B}+2\theta_{C}}
(A\cup B)(CD)^{2}\partial_{A}\left( AB\right)^{2}.
\end{eqnarray*}
In going between the first and second line here we have assumed $f$ is either
even or odd. If either $A$ or $B$ is empty the result is understood to
be vanishing.
Similarly
\begin{align*}
\left\{ e^{\theta _{I}}(I),e^{-\theta _{J}}(J)\right\} & =
e^{\theta_{I}-\theta_{J}}\left[ (I)\partial_{I}(J)+(J)\partial_{J}(I)\right]\\
& =(\pm)\sp{(|A|+|B|)|C|+|A||B|}
e^{\theta_{I}-\theta_{J}}\left( AB\right)\sp2 \partial _{C}(CD)^{2} .
\end{align*}
Now $\{S_k,S_l\}$ will vanish if and only if the coefficients of the
independent $e^{\theta_{I}+\theta _{J}}=e^{\theta_{A\cup B}+2\theta_{C}}$
vanish. Such a term fixes the sets $C$, $D$ and $A\cup B$, and the same
momentum dependence appears for each subset $L\subseteq A\cup B$
with $|L|=|A|$.
The coefficient of $e^{\theta_{A\cup B}+2\theta_{C}}$ in $\{S_k,S_l\}$
then vanishes provided
$$\sum\limits_{\substack{ L\subseteq A\cup B \\ |L|=|A|}}\partial _{L}
(L\, A\cup B \backslash L)\sp2=0.
$$
Therefore (with $k,l\in \{1,2,\ldots, n\}$)
\begin{equation}
\{S_k,S_l\}=0
\Longleftrightarrow
\sum\limits_{\substack{ L\subseteq \{1,2,\ldots, k+l-2c\} \\ |L|=k-c}}
\partial _{L}(L\, \{1,2,\ldots, k+l-2c\}\backslash L)\sp2=0,
\label{gencons}
\end{equation}
for all $c$ satisfying $\max(k+l-n,0)\leq c\leq \min(k,l)$.
In particular
\begin{equation}
\{S_1,S_l\}=0 \Longleftrightarrow
\sum_{j=1}\sp{l+1}\partial_j \prod_{k\ne j}f\sp2(x_j-x_k)=0
\label{functionalm}
\end{equation}
and we recover (\ref{functional}) for $l=n-1$.

If we take $k\le l\le n-1$ the term $c=k-1$ is allowed in the sum leading to
\begin{equation}
\{S_k,S_l\}=0 \Longrightarrow
\sum_{j=1}\sp{l-k+2}\partial_j \prod_{s\ne j}\sp{l-k+2}f\sp2(x_j-x_s)=0
\end{equation}
We know our new functions do not satisfy this when $l-k$ is odd. Thus the
conserved quantities of Ruijsenaars will not Poisson commute
for this new model. We cannot as yet rule out other Poisson commuting
conserved quantities. In the case of $n=4$ we note that $S_1$, $S_3$ and
$S_4$ Poisson commute.

\section{Scaling Limits}
We will now construct some classical Hamiltonian systems from our new solutions.
We begin by recalling the scaling limit of the Ruijsenaars-Schneider model
\cite{RS} that results in the Calogero-Moser model.
Whereas the Calogero-Moser models are (completely integrable)
natural Hamiltonian systems, the scaling limits we must consider
have a different form. There are also analogous systems corresponding to the
these scaling limits of the Ruijsenaars-Schneider models.

Let us write $p_j={\bar p}_j/c$ and $x_j= c {\bar x}_j$, a scaling
which preserves the Poisson brackets
$\{x_i,p_j\}=\delta_{ij}=\{{\bar x}_i,{\bar p}_j\}$.
Putting in dimensionful parameters we have
\begin{equation*}
\begin{split}
\lim_{c\rightarrow\infty}(H-nmc\sp2)&=
\lim_{c\rightarrow\infty}\Bigg(
m c\sp2\sum_{j=1}\sp{n}\cosh \frac{{\bar p}_j}{mc}\, \prod_{k\ne j}
\sqrt{1+\frac{g^2}{c^2} \wp(\frac{{\bar x}_j-{\bar x}_k}{\bar A}) }
-nmc\sp2\Bigg)\\
&= \frac{1}{2}\sum_{j=1}\sp{n} \frac{{\bar p}_j\sp2}{m}+
\frac{m g\sp2}{2}\sum_{j\ne k} \wp(\frac{{\bar x}_j-{\bar x}_k}{\bar A})
\end{split}
\end{equation*}
Here the parameter $A$ has been introduced \cite{BS} to make the argument of
the $\wp$ function dimensionless: $\wp(x/A)=\wp({\bar x}/{\bar A})$, thus
$A$ scales as $c {\bar A}$.
This scaling limit is possible because of the (assumed) nonzero constant
that can appear in the Ruijsenaars-Schneider solution (here scaled to one).
Such a constant is not present in our new solutions, and the scaling does not
work if the constant in the Ruijsenaars-Schneider solution is zero. We shall
now consider a different scaling.

Both the Ruijsenaars systems and ours have a different type of
scaling limit
$$
\lim_{c\rightarrow\infty}\Bigg(
m c\sp2\sum_{j=1}\sp{n}\cosh \frac{{\bar p}_j}{mc}\, \prod_{k\ne j}
f(\frac{{\bar x}_j-{\bar x}_k}{\bar A})-\lambda m c\sp2\Bigg)=
\frac{1}{2}\sum_{j=1}\sp{n} \frac{{\bar p}_j\sp2}{m}
\prod_{k\ne j}
f(\frac{{\bar x}_j-{\bar x}_k}{\bar A})
$$
Here we have introduced
\begin{equation}
\lambda=\sum_{j=1}\sp{n}\,\prod_{k\ne j}f(\frac{{\bar x}_j-{\bar x}_k}{\bar A}),
\label{lamdef}
\end{equation}
which need not be constant. In some special cases however it is.
We know for example that if $\lambda$ is a constant, $f(x)$
is an odd function, and
\begin{enumerate}
\item (\ref{lamdef}) is true for all $n\ge3$ then
       $f(x)=1/x$ ($\lambda=0$), $\coth(x)$ ($\lambda=1$ for $n$ odd
        and $\lambda=0$ for $n$ even),
\item  (\ref{lamdef}) is true for all even $n\ge4$ then
       $f(x)=\sqrt{\wp(x)-e_\alpha}$ ($\lambda=0$).
\end{enumerate}
This type of degeneration of the Ruijsenaars-Schneider models does not
appear to have been considered before. (The same will work with the addition
of extra potentials not considered here.) The functional identities
being used here also underly the various Hirzebruch genera \cite{HBJ}.
We find for example that
\begin{thm} The functions
$$
H=\frac{1}{2}\sum_{j=1}\sp{n}{p}_j\sp2\prod_{k\ne j} f(x_j-x_k),\qquad
P=\sum_{j=1}\sp{n}{p}_j\prod_{k\ne j} f(x_j-x_k),\qquad
B=\sum_{j=1}\sp{n}x_j 
$$
obey the algebra
\begin{equation}
\{H,B\}=P ,\qquad \{P,B\}=\lambda ,\qquad \{H,P\}=0.
\label{galilei}
\end{equation}
if and only $f(x)$ is either an even or odd function satisfying
$$
\sum_{j=1}\sp{n}\,\prod_{k\ne j}f( x_j-x_k)=\lambda,
$$
where $\lambda$ is a constant. In particular, the odd functions
$f(x)=1/x$ ($\lambda=0$), $\coth(x)$ ($\lambda=1$ for $n$ odd
        and $\lambda=0$ for $n$ even),
$\sqrt{\wp(x)-e_\alpha}$ ($\lambda=0$) yield solutions.
\end{thm}
When $\lambda=0$ this is the Galilean algebra, while $\lambda\ne0$
is a central extension of the Galilean algebra.
Our new functions do not satisfy this unless we have a degeneration.
Interestingly, in the case of an even number of particles, particular
cases of the elliptic Ruijsenaars-Schneider model are in this list.
Whatever, we can ask whether the metrics this scaling limit yield
have any special significance.
We are encountering models of the form
$ H=\frac{1}{2}\sum_{j=1}\sp{n} g\sp{jj} p_j\sp2$ and so dealing with
diagonal metrics.
We cannot as yet characterise these metrics, but towards this report
\begin{thm} The diagonal metric
\begin{equation}
ds^2=\sum_{i=1}^n \Big(\prod_{j\ne i}\,\Psi(x\sp{i}-x\sp{j})\Big) (dx^i)^2,
\label{diagm}
\end{equation}
with potentially nonvanishing curvature components $R\sp{i}_{\, jik}$,
$R\sp{i}_{\, jjk}$ ($k\ne i,j$) and $ R\sp{i}_{\, jij}$ has
\begin{enumerate}
\item $R\sp{i}_{\, jik}=R\sp{i}_{\, jjk}=0$ ($k\ne i,j$) if and only if
$\Psi(x)=\alpha \big( e\sp{2 bx}-1\big)\sp{a}$ or $\alpha x\sp{a}$.
We may set $\alpha=1$ by rescaling $x$.
\item $R\sp{i}_{\, jij}= (-1)\sp{n}\, b\sp2$ when
$\Psi(x)=\big( e\sp{2 bx}-1\big)$,
\item  $R\sp{i}_{\, jij}=0$ when $\Psi(x)= x$.
\end{enumerate}
Thus $\Psi(x)= x$ yields a solution of the Lam\'e equations.
\end{thm}
These metrics are of St\"ackel form. The rational degenerations of our
models are given by this theorem. They may be understood as a  parabolic
limit of Jacobi elliptic coordinates\footnote{We thank E. Ferapontov for
discussion on this matter.}.
Indeed, although we have investigated metrics of the form (\ref{diagm})
because these arise from our systems, we have more generally
\begin{col}
The diagonal metric
\begin{equation}
ds^2=\sum_{i=1}^n \Big(\chi_i(x\sp{i})\,
\prod_{j\ne i}\,\Psi(x\sp{i}-x\sp{j})\Big) (dx^i)^2,
\label{diagm2}
\end{equation}
has $R\sp{i}_{\, jik}=R\sp{i}_{\, jjk}=0$ ($k\ne i,j$) if and only if
$\Psi(x)=\alpha \big( e\sp{2 bx}-1\big)\sp{a}$ or $\alpha x\sp{a}$.
\end{col}

\section{Quantum Models}
Ruijsenaars \cite{R} also investigated the quantum version of the classical
models he and Schneider introduced. From the outset he sought operator
analogues of the light-cone quantities (\ref{lccons}).
He showed that (for $k=1,\ldots n$)
\begin{equation*}
{\hat S}_{k} =\sum_{\substack{ I\subseteq \{1,2,\ldots, n\}\\ |I|=k} }
\prod_{\substack{ i\in I \\ {j\not\in I}}} h(x_j-x_i)\sp{\frac{1}{2}}\,
\exp\left(-\sqrt{-1}\,\beta {\sum_{i\in I} \partial_i}\right) \,
\prod_{\substack{ i\in I \\ j\not\in I}} h(x_i-x_j)\sp{\frac{1}{2}}
\label{qlccons}
\end{equation*}
pairwise commute if and only if (for all $k$ and $n\ge1$)
\begin{equation}
\sum\limits_{\substack{ I\subseteq \{1,2,\ldots, n\} \\ |I|=k}}
\Bigg( \prod_{\substack{ i\in I \\ {j\not\in I}}} h(x_j-x_i)
h(x_i-x_j-i\beta)-
\prod_{\substack{ i\in I \\ {j\not\in I} }} h(x_i-x_j)
h(x_j-x_i-i\beta)\Bigg)=0.
\label{qgencons}
\end{equation}
Here $\beta$ is an arbitrary positive number. Upon dividing
by $\beta$ and letting $\beta\rightarrow0$ this yields (\ref{gencons})
and in particular (\ref{functional}) with $F(x)=h(x)h(-x)$.
(Given $F(x)$ there is of course an an ambiguity in $h(x)$ under
$h(x)\rightarrow e\sp{bx}h(x)$.)
Ruijsenaars was able to show that $h(x)=\sigma(x+\mu)/(\sigma(x) \sigma(\mu))$
led to a solution of (\ref{qgencons}), the solution being related to the
earlier Ruijsenaars-Schneider solution via
$$\frac{\sigma(x+\mu)\sigma(x-\mu)}{\sigma\sp2(x) \sigma\sp2(\mu)}=
\wp(\mu)-\wp(x).
$$
Ruijsenaars \cite{R} suggested that this solution was ``most likely unique"
but was unable to prove this.
A consequence of our classical analysis are the possible functions
$F(x)=h(x)h(-x)$.
A natural question to ask is whether there is a solution to (\ref{qgencons})
corresponding to our new solutions. If not, then the
Ruijsenaars solution is indeed unique.
At present we are able to show that the Ruijsenaars solution is the unique
solution to (\ref{qgencons}) with $k=1$ for $n=3,4$ while  for general $n$ 
and those $h(x)$ with infinite real period and tending to $0$ at infinity, 
the only solutions correspond to $F(x)=1/\sinh\sp2 x$ and $F(x)=1/x^2$.
In particular there is no $h(x)$ corresponding to our new solution
$F(x)=\cosh(x)/\sinh\sp2 x$.
We anticipate being able to extend this to show
that none of the new solutions can be written in
the form $F(x)=h(x)h(-x)$ with $h(x)$ satisfying (\ref{qgencons}).

\section{Conclusion}
In this letter we have investigated  mechanical systems exhibiting an
action of the Poinc\'are algebra. By adopting a given ans\"atz this reduces to
a study of the functional equation (\ref{functional}) and we presented the
general solution to this. Beyond the solution of Ruijsenaars and Schneider new
solutions are obtained. The conserved quantities of Ruijsenaars and Schneider
do not Poisson commute for these new models. Although we cannot as yet
rule out other Poisson commuting conserved quantities, our models suggest
that Poinc\'are invariance and integrability are distinct requirements.
A new classical limit of our models (and the Ruijsenaars-Schneider model)
was discussed. Here a diagonal metric of a particular type arises.
The rational degenerations of these models satisfy the Lam\'e equations
and indeed are the only ones within our class of metrics that do.
Finally we have considered the quantum mechanical version of our new models.
While not complete, we can show for $n=3,4$ that the only solutions to
Ruijsenaars functional equations (\ref{qgencons}) are those given by
Ruijsenaars and Schneider, and that for general $n$ the degenerations of our 
new models, where distinct from the Ruijsenaars solutions, fail to yield a 
solution. Our new techniques suggest
a proof of the uniqueness of the  Ruijsenaars solution is now within reach,
and this will be pursued elsewhere.

\section{Acknowledgements}
We wish to thank A. M. Davie for helpful discussion and S.N.M. Ruijsenaars
for critically reading the manuscript.
Aspects of this work have been described at the BAMC (Reading:April, 2001), the
mini-workshop ``The  Calogero-Moser System  Thirty Years Later"
(Rome: May, 2001), NEEDS (Cambridge: July, 2001) and the EuroWorkshop
``Discrete systems and integrability" (Cambridge: September, 2001).
We are grateful to the organisers of these meetings and to the participants
for their suggestions.
One of the authors (H.W.B.) wishes to thank the Newton Institute for
support during the completion of this work.

\end{document}